\documentclass[aps,prc,twocolumn,superscriptaddress,showpacs]{revtex4}
\usepackage{graphicx,tabularx,dcolumn,amsmath}
\begin{document}


\title{
Low-energy cross section of the $^7$Be(p,$\gamma$)$^8$B solar
fusion reaction from Coulomb dissociation of $^8$B}

\author{F.~{Sch{\"u}mann}}
\affiliation{Institut f{\"u}r Physik mit Ionenstrahlen,
Ruhr-Universit{\"a}t Bochum, D-44780 Bochum, Germany}
\author{S.~{Typel}}
\affiliation{Gesellschaft f{\"u}r Schwerionenforschung, D-64220
Darmstadt, Germany}
\author{F.~{Hammache}}
\altaffiliation[Present address: ]{IPN Orsay, F-91406 Orsay Cedex,
France} \affiliation{Gesellschaft f{\"u}r Schwerionenforschung,
D-64220 Darmstadt, Germany}
\author{K.~{S{\"u}mmerer}}
\email[Corresponding author: ]{k.suemmerer@gsi.de}
\affiliation{Gesellschaft f{\"u}r Schwerionenforschung, D-64220
Darmstadt, Germany}
\author{F.~{Uhlig}}
\affiliation{Gesellschaft f{\"u}r Schwerionenforschung, D-64220
Darmstadt, Germany}
\affiliation{Technische Universit{\"a}t
Darmstadt, D-64289 Darmstadt, Germany}
\author{I.~{B{\"o}ttcher}}
\affiliation{Phillips-Universit{\"a}t Marburg, D-35032 Marburg,
Germany}
\author{D.~{Cortina}}
\altaffiliation[Present address: ]{Universidad Santiago de
Compostela, Spain} \affiliation{Gesellschaft f{\"u}r
Schwerionenforschung, D-64220 Darmstadt, Germany}
\author{A.~{F{\"o}rster}}
\altaffiliation[Present address: ]{CERN, CH-12211 Geneve 23,
Switzerland} \affiliation{Technische Universit{\"a}t Darmstadt,
D-64289 Darmstadt, Germany}
\author{M.~{Gai}}
\affiliation{University of Connecticut, Groton, CT 06340, U.S.A.}
\author{H.~{Geissel}}
\affiliation{Gesellschaft f{\"u}r Schwerionenforschung, D-64220
Darmstadt, Germany}
\author{U.~{Greife}}
\affiliation{Department of Physics, Colorado School of Mines,
Golden, CO 80401, U.S.A.}
\author{E.~{Grosse}}
\affiliation{Forschungszentrum Rossendorf, D-01314 Dresden,
Germany}
\author{N.~{Iwasa}}
\affiliation{Tohoku University, Aoba, Sendai, Miyagi 980-08578,
Japan}
\author{P.~{Koczo{\'n}}}
\affiliation{Gesellschaft f{\"u}r Schwerionenforschung, D-64220
Darmstadt, Germany}
\author{B.~{Kohlmeyer}}
\affiliation{Phillips-Universit{\"a}t Marburg, D-35032 Marburg,
Germany}
\author{R.~{Kulessa}}
\affiliation{Jagellonian University, PL-30-059 Krak{\'o}w, Poland}
\author{H.~{Kumagai}}
\affiliation{RIKEN (Institute of Physical and Chemical Research),
Wako, Saitama 351-0198, Japan}
\author{N.~{Kurz}}
\affiliation{Gesellschaft f{\"u}r Schwerionenforschung, D-64220
Darmstadt, Germany}
\author{M.~{Menzel}}
\affiliation{Phillips-Universit{\"a}t Marburg, D-35032 Marburg,
Germany}
\author{T.~{Motobayashi}}
\affiliation{Rikkyo University, Toshima, Tokyo 171, Japan}
\affiliation{RIKEN (Institute of Physical and Chemical Research),
Wako, Saitama 351-0198, Japan}
\author{H.~{Oeschler}}
\affiliation{Technische Universit{\"a}t Darmstadt, D-64289
Darmstadt, Germany}
\author{A.~{Ozawa}}
\altaffiliation[Present address: ]{University of Tsukuba, Ibaraki,
Japan} \affiliation{RIKEN (Institute of Physical and Chemical
Research), Wako, Saitama 351-0198, Japan}
\author{M.~{P{\l}osko{\'n}}}
\affiliation{Jagellonian University, PL-30-059 Krak{\'o}w, Poland}
\author{W.~{Prokopowicz}}
\affiliation{Jagellonian University, PL-30-059 Krak{\'o}w, Poland}
\author{E.~{Schwab}}
\affiliation{Gesellschaft f{\"u}r Schwerionenforschung, D-64220
Darmstadt, Germany}
\author{P.~{Senger}}
\affiliation{Gesellschaft f{\"u}r Schwerionenforschung, D-64220
Darmstadt, Germany}
\author{F.~Strieder}
\affiliation{Institut f{\"u}r Physik mit Ionenstrahlen,
Ruhr-Universit{\"a}t Bochum, D-44780 Bochum, Germany}
\author{C.~{Sturm}}
\affiliation{Gesellschaft f{\"u}r Schwerionenforschung, D-64220
Darmstadt, Germany}
\affiliation{Technische Universit{\"a}t
Darmstadt, D-64289 Darmstadt, Germany}
\author{Zhi-Yu~{Sun}}
\altaffiliation[Present address: ]{Inst. of Modern Physics,
Lanzhou, China} \affiliation{Gesellschaft f{\"u}r
Schwerionenforschung, D-64220 Darmstadt, Germany}
\author{G.~{Sur{\'o}wka}}
\affiliation{Gesellschaft f{\"u}r Schwerionenforschung, D-64220
Darmstadt, Germany}
\affiliation{Jagellonian University, PL-30-059
Krak{\'o}w, Poland}
\author{A.~{Wagner}}
\affiliation{Forschungszentrum Rossendorf, D-01314 Dresden,
Germany}
\author{W.~{Walu{\'s}}}
\affiliation{Jagellonian University, PL-30-059 Krak{\'o}w, Poland}

\date{\today}

\begin{abstract}
An exclusive measurement of the Coulomb breakup of $^8$B into
$^7$Be+p at 254 $A$ MeV was used to infer the low-energy
$^7$Be(p,$\gamma$)$^8$B cross section. The radioactive $^8$B beam
was produced by projectile fragmentation of 350 $A$ MeV $^{12}$C
and separated with the fragment separator FRS at GSI in Darmstadt,
Germany. The Coulomb-breakup products were momentum-analyzed in
the KaoS magnetic spectrometer; particular emphasis was placed on
the angular correlations of the breakup particles. These
correlations demonstrate clearly that E1 multipolarity dominates
within the angular cuts selected for the analysis. The deduced
astrophysical $S_{17}$ factors exhibit good agreement with the
most recent direct $^7$Be(p,$\gamma$)$^8$B measurements. By using
the energy dependence of $S_{17}$ according to the recently
refined cluster model for $^8$B of Descouvemont, we extract a
zero-energy $S$ factor of $S_{17}(0) = 20.6 \pm 0.8 (stat) \pm 1.2
(syst)$ eV b. These errors do not include the uncertainty of the
theoretical model to extrapolate to zero relative energy,
estimated to be about 5\% by Descouvemont.
\end{abstract}

\pacs{25.40.Lw, 25.60.-t, 25.70.De, 26.65.+t}

\maketitle

\section{Introduction}

The so-called ``solar neutrino problem'' has been solved by the
results of the Sudbury Neutrino Observatory
(SNO)~\cite{Ahm01,SNO05}. The SNO experiment has shown strong
evidence that the neutrino-flux deficit measured in
charged-current interactions is a result of neutrino flavor
oscillations between electron-neutrino production in the Sun and
their detection on Earth. The flux measured in neutral-current
interactions of high-energy solar neutrinos is in general
agreement with the flux predicted by the standard solar model
(SSM, Refs.~\cite{BP04,Tur04}). The current slight discrepancy
between the flux predicted by the SSM and the neutral-current flux
measured by SNO~\cite{SNO05} may be significant or not depending
on the uncertainty of the flux prediction: a small but significant
deficit could, e.g., be evidence for oscillations into sterile
neutrinos. To that end it is essential to further reduce the
uncertainty of nuclear inputs to the SSM in order to refine its
predictions. One critical quantity is the $^7$Be(p,$\gamma$)$^8$B
cross section at solar energies since it is linearly related with
the high-energy solar neutrino flux stemming from $^8$B
$\beta$-decay.

In recent years, many attempts have been undertaken to measure
this cross section with high-precision in direct-proton-capture
measurements using radioactive $^7$Be targets
~\cite{Ham01,Str01,Bab03,Jun03}. Unfortunately, these results do
not yield a completely consistent picture: The earlier
measurements (Refs.~\cite{Ham01,Str01}) yield lower zero-energy
astrophysical $S$ factors, $S_{17}(0)$, around 19 eV b, whereas
the two more recent ones (Refs.~\cite{Bab03,Jun03}) obtain results
which are about 15\% higher. All (p,$\gamma$) data sets, however,
were found to be consistent with an energy dependence of $S_{17}$
as given by the cluster model of Descouvemont and Baye
~\cite{Des94}.

In view of their importance for astro- and elementary-particle
physics, it is desirable to cross-check these results by other,
indirect measurements that have different systematic errors. One
possibility is Coulomb dissociation (CD) of $^8$B in the
electromagnetic field of a high-$Z$ nucleus. Such measurements
have been performed at intermediate \cite{Kik98,Dav01} and high
energies \cite{Iwa99}. The present paper reports on a CD
experiment similar to that of Iwasa \textit{et al.} (GSI-1, Ref.
\cite{Iwa99}), but with an improved experimental technique: In
GSI-1, the incident $^8$B beam could not be tracked before the
target, whereas in the present run we could measure the angles
before and after the target with good precision. Preliminary
results of the present study have been published
earlier~\cite{Sch03}. For the present publication, the data were
re-analyzed, leading to slightly different results for the
lowest-energy data points.

Another indirect method to deduce $S_{17}(0)$ is to determine the
Asymptotic Normalization Coefficients (ANC) of the proton wave
functions bound in the $^7$Be potential. This method makes use of
the fact that due to the very low proton binding energy radiative
proton capture is extremely peripheral and $S_{17}(0)$ can be
calculated  directly from the ANC. These ANC are determined from
low-energy proton-transfer or from proton-removal cross sections
\cite{Muk01,Tra04,Cor03}. A recent re-examination of ANC-results
for $^7$Be(p,$\gamma$) by Trache \textit{et al.} yielded a
relatively small central value of $S_{17}(0)= 18.7 \pm 1.9$ eV
b~\cite{Tra04}. Still, this value is in line with all published
values of $S_{17}(0)$ except for Ref.\cite{Jun03}.

It is important to compare the results from direct and indirect
methods to determine the astrophysical $S$-factors with each other
since the indirect methods could also be used to study
astrophysically interesting reactions between {\em unstable}
nuclei where direct-capture reactions cannot be applied. The
reaction $^7$Be(p,$\gamma$)$^8$B could be an ideal test case,
provided that the remaining inconsistencies in $S_{17}$(0) from
the different, direct and indirect, methods can be resolved.

In the present paper, we give a comprehensive report of the CD
experiment performed at GSI~\cite{Sch03} where we impinged a
secondary $^8$B beam with an incident energy of 254 $A$ MeV on a
$^{208}$Pb break up target. As already mentioned in our earlier
publication, we focus on a crucial question that must be answered
if one wants to use the CD method to derive a precise value for
$S_{17}$(0): the contribution of E2 multipolarity to CD of $^8$B.
One can calculate that E1 is the dominant multipolarity in CD as
well as in direct proton capture, but it is obvious that the
equivalent photon field from a high-$Z$ target nucleus seen by a
projectile at a few hundred MeV per nucleon contains also a strong
E2 component. Experimental limits for a possible E2 contribution
were extracted in the work of Kikuchi \textit{et al.}~\cite{Kik98}
and Iwasa \textit{et al.}~\cite{Iwa99}; both papers found
negligible E2 contributions. Recently, Davids \textit{et al.} have
reported positive experimental evidence for a finite E2
contribution in CD of $^8$B, mainly from the analysis of {\em
inclusive} longitudinal momentum ($p_{||}$) spectra of $^7$Be
fragments measured at 44 and 81 $A$ MeV \cite{Dav01,Dav03}; they
therefore subtracted a calculated E2 contribution from their
$S_{17}$ data. In order to resolve these discrepancies, we have
analyzed observables that should be particularly sensitive to
contributions from E2 multipolarity, namely the angular
correlations of the $^8$B-breakup particles, proton and $^7$Be.

\section{Model calculations}

Accurate model calculations of the CD of $^8$B are essential for
several reasons. From a practical point of view, the relatively
bad energy resolution of the CD method requires to simulate e.g.
the effects of cross talk between neighboring energy bins, of the
finite size and resolution of the tracking detectors etc. These
simulations require a CD event generator that is reasonably close
to reality so that the remaining differences between the measured
and simulated cross-section distributions can be attributed to the
$S_{17}$ factor. For this purpose we have used a simple potential
model of $^8$B.

Since the current experiment (like most other direct and indirect
studies of the $^7$Be(p,$\gamma$)$^8$B reaction) does not allow to
measure at solar energies, the data set has to be extrapolated
towards $E_{rel}=0$. For this purpose we have to use the most
sophisticated model available. We will show below that a cluster
model of $^8$B~\cite{Des04} seems to be suited best for a reliable
extrapolation.

\subsection{Nuclear-structure of $^8$B}

\begin{figure}[bt]
\includegraphics[width=7cm]{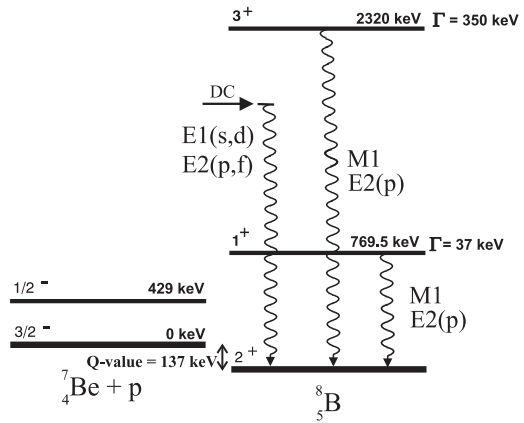}
\caption{Level schemes of $^7$Be and $^8$B relevant for direct
proton capture into the ground state of $^8$B and for resonant
capture via the M1 resonances at 0.770 MeV and 2.32 MeV excitation
energy.} \label{level_scheme}
\end{figure}

The isotope $^8$B has one of the lowest proton binding energies of
all particle-stable nuclei known in the chart of nuclides ($B_p=
137.5$ keV, \cite{Til04}). The relevant parts of the $^7$Be and
$^8$B level schemes are depicted in Fig.\ref{level_scheme}.
The simplest model for $^8$B is that of a p-wave proton coupled to
an inert $^7$Be core with $I^{\pi}=3/2^{-}$ to form the $^8$B
ground state with $I^{\pi}=2^{+}$. We have adopted this simplified
single-particle model of $^8$B to calculate cross sections within
the simulations to be described below. Details of the model are
described in Ref.~\cite{Dav03}. The proton is bound in a
Woods-Saxon potential with radius parameter $r_0 = 1.25$ fm and
diffuseness $a = 0.65$ fm. As usual, the p-wave potential depth
has been adjusted to match the $^8$B proton binding energy, this
yields a depth of 43.183 MeV. The s-, d- and f-wave potentials
have been adjusted to reproduce the s-wave scattering lengths of
the mirror $^7$Li+n reaction \cite{Koe83}, this yields $V_1 =
43.857$ MeV for channel spin $S$=1 and $V_2 = 52.597$ MeV for
channel spin $S$=2. We note that we obtain for the dominant
channel spin $S$=2 an s-wave scattering length for $^7$Be+p of
$a_{02}^{theo} = -8$ fm which agrees well with the recently
measured value of $a_{02}^{exp} = -7 \pm 3$ fm (Angulo \textit{et
al.}~\cite{Ang03}).

\begin{figure}[bt]
\hspace{-1.0cm}
\includegraphics[width=9.0cm]{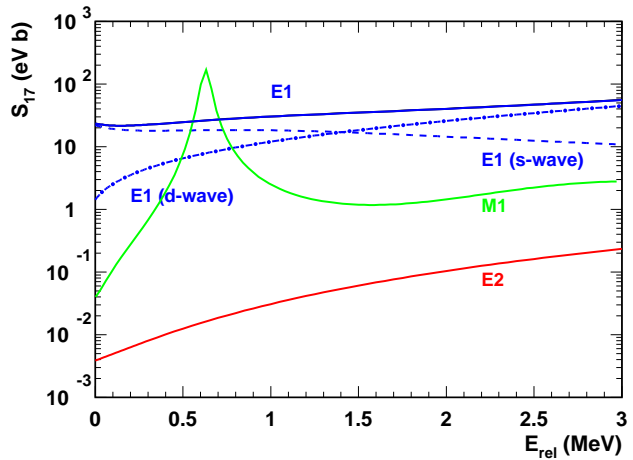}
\caption{(Color online)
Theoretical $S_{17}$ factors from a simple potential model of $^8$B
and their decomposition
into contributions from various partial waves.
}
\label{s17_pg_theo}
\end{figure}

With this model we obtain astrophysical $S$-factors as a function
of the proton-$^7$Be relative energy, $E_{rel}$, as shown in
Fig.\ref{s17_pg_theo}. The non-resonant direct capture into the
$^8$B ground state proceeds mainly via s- and d-wave captures and
E1 $\gamma$-emission. Capture of p- and f-waves followed by E2
emission plays an insignificant role, in particular at solar
energies. The resonant component proceeds through the $1^+$
resonance at 770 kev (632 keV above threshold) which decays mainly
by M1 emission and is limited essentially to a narrow region
around the resonance energy, with minor but finite contributions
at relative energies above the resonance. The M1-resonance cross
section has been obtained from experimental data~\cite{Fil83}, it
is not contained in the model of Ref.~\cite{Dav03}. We have
ignored entirely the high-lying M1 resonance at 2.32 MeV since it
cannot be seen in our high-energy CD experiment, due to both, the
small cross section in CD and its large width.

The potential model of $^8$B sketched above ignores the well known
cluster structure of $^7$Be (see e.g. Ref.~\cite{Nun97}).
Descouvemont and Baye~\cite{Des94} have therefore applied a model
where $^8$B is assumed to consist of e.g. p+($^7$Be=$^3$He+$^4$He)
or $^3$He+($^5$Li=p+$^4$He) three-cluster structures, including
excited states of the clusters. Later, this model was slightly
refined~\cite{Des04} by allowing also for variations in the
cluster separation and by using different effective
nucleon-nucleon interactions; the results were found to be largely
unchanged. Our potential model can be viewed as a simplification
of the cluster model where the cluster distance is artificially
set to zero; the spectroscopic information available for $^7$Be
points, however, to a finite cluster distance of $\approx$3.5
fm~\cite{Des04}.

For computational simplicity, we will use the potential model to
simulate the differential observables in our experiment and come
back to the cluster model at the end of this article where we
discuss the extrapolation of $S_{17}$ towards zero relative
energy.

\subsection{Coulomb dissociation of $^8$B}

As proposed by Baur \textit{et al.} \cite{Bau86}, CD can be used
favorably to measure radiative-capture cross sections by making
use of the strong flux of equivalent photons originating from a
heavy target nucleus as seen by a fast-moving projectile, which
replaces the presently insufficient intensity of available
real-photon sources. Assuming first-order perturbation theory for
the electromagnetic excitation process, CD cross sections can be
converted directly to photo-dissociation cross sections. The
latter are related to the astrophysically relevant
radiative-capture cross sections by the principle of detailed
balance. It is obvious that the indirect method of CD needs
theoretical input in the conversion process.

There are several sources for complications. They can be
identified and minimized by selecting appropriate observables and
kinematical conditions in the experiment.
\begin{enumerate}
\item Several multipolarities (E1, E2, M1, \dots) with different
weights contribute in radiative capture reactions and Coulomb
breakup. In principle, they can be disentangled by studying
angular distributions in CD, preferably in the center-of-mass
system of the excited nucleus. Their relative strengths depend on
projectile energy and scattering angle. \item An exchange of more
than one photon (``higher-order effects'') destroys the direct
relation between the CD cross section and the photo-dissociation
cross section. High projectile energies and large impact
parameters reduce this effect. \item The nuclear interaction
between projectile and target induces nuclear breakup and
absorption. It becomes relevant for small impact parameters.
\end{enumerate}
These are general features of CD that can be included in the
theoretical description of the breakup mechanism in various
approximations which lead to corrections of the simple
pure-Coulomb first-order approach.

\subsubsection{Semiclassical calculations}

The Coulomb-breakup mechanism can be described both in fully
quantal approaches and semi-classical models \cite{Ald56}. In the
latter case the projectile is assumed to move on a classical
trajectory with respect to the target. In our case, we use the
semi-classical model in first-order perturbation theory (PT) to
describe the CD of $^8$B in the Coulomb field of $^{208}$Pb, as
described in more detail elsewhere \cite{Ber94,Typ02,Dav03}. The
excitation amplitude is calculated in the relativistic approach
assuming a straight-line trajectory but correcting the excitation
functions for the deflection in the Coulomb field of the target
\cite{Win79}. This is appropriate at the high incident energy used
in the present experiment (254 $A$ MeV) and justified {\it a
posteriori} by the good agreement with the measured angular
distributions.

In addition to CD, nuclear overlap of $^8$B and $^{208}$Pb has to
be considered. This will mainly take flux out of the $^7$Be+p exit
channel; feeding this channel by nuclear interaction has been
calculated to be of minor importance by Bertulani and Gai
\cite{Ber98}. In order to take nuclear absorption into account we
modified the relativistic Coulomb-excitation functions by
multiplying them with a correction factor as described in Ref.
\cite{Typ02}. This factor is derived from an eikonal approximation
of the excitation functions taking both Coulomb and nuclear
potentials into account. In the present case we assume a diffuse
absorptive nuclear potential with a depth of 20 MeV and a radius
of 9.91 fm, i.e.\ the sum of the projectile and target radii. As
we will see below, this choice reproduces well the integral
scattering-angle distribution.

\subsubsection{Dynamical calculations}
Higher-order effects from the exchange of more than one photon can
be considered in semi-classical calculations that study the
time-evolution of the projectile system during the scattering. As
compared to a first-order calculation, the momenta of the outgoing
particles are modified in the Coulomb field of the target leading
to a distortion of relative-energy and angular-momentum
distributions. Esbensen \textit{et al.}~\cite{Esb04,Esb05} have
proposed that discrepancies between the results from
radiative-capture and from CD studies of the
$^7$Be(p,$\gamma$)$^8$B reaction are due to deficiencies of the
method how to evaluate $S_{17}$ from CD cross sections by using
first-order PT. They point out that a full dynamical calculation
of CD, if compared to a first-order PT calculation, will lead to
an increased $S_{17}$ factor at low $E_{rel}$ and a reduced one at
high $E_{rel}$, thus producing a smaller slope of $S_{17}$ {\it
vs.} $E_{rel}$ and a better agreement between the results from the
two methods. However, the amount of this modification depends on
the assumed E2 strength and thus is model dependent. Recently,
fully quantal calculations became available that consider the
post-acceleration of the fragments in the Coulomb field of the
target. In contrast to dynamical calculations in the semiclassical
approach, they predict an {\em increase} of the cross section at
low relative energies \cite{Alt05}.  More theoretical work is
required in order to obtain a consistent picture of higher-order
effects.

To follow the suggestions of Esbensen \textit{et al.}, we have
also performed dynamical calculations of the CD of $^8$B at 254
$A$ MeV following the approach as described in \cite{Dav03} for
lower projectile energies assuming the simple potential model for
${}^{8}$B.  The ${}^{8}$B nucleus moves on a Coulomb trajectory
taking the deflection into account. E1 and E2 multipoles were
considered in the standard far-field approximation with the full
strength as predicted by the model.

In both theoretical approaches, triple-differential cross sections
for the CD of ${}^{8}$B are obtained. These distributions of
observables cannot be compared directly to the measured data, but
have to be folded with the respective experimental resolutions. To
this end, the cross sections were converted to statistically
distributed ``event'' distributions from both (PT and dynamical)
calculations and run through our experimental filter as will be
described in more detail below.

\section{Experimental procedures}

Several other CD studies of $^8$B breakup \cite{Kik98,Dav01} have
used intermediate energies between 46 and 83 $A$ MeV as available
from cyclotron-based radioactive-beam facilities. At GSI, the 18
Tm SIS-18 synchrotron allows to go much higher in incident energy.
We have chosen a $^8$B incident energy of 254 $A$ MeV limited by
the maximum bending power of the KaoS spectrometer used for
determining the momenta of the break-up particles, p and $^7$Be.
In the following we will describe in detail the preparation and
identification of the secondary beam as well as the experimental
equipment used to measure the breakup.

\subsection{Preparation and properties of the $^8$B beam}

The $^8$B secondary beam was produced at the SIS/FRS
radioactive-beam facility at GSI~\cite{Gei92} by fragmenting a 350
$A$ MeV $^{12}$C beam in a 8 g/cm$^2$ Be target and separating it
from contaminant ions in a 1.4 g/cm$^2$ wedge-shaped Al degrader
placed in the FRS intermediate focal plane.

\begin{figure}[bt]
\includegraphics[width=9cm]{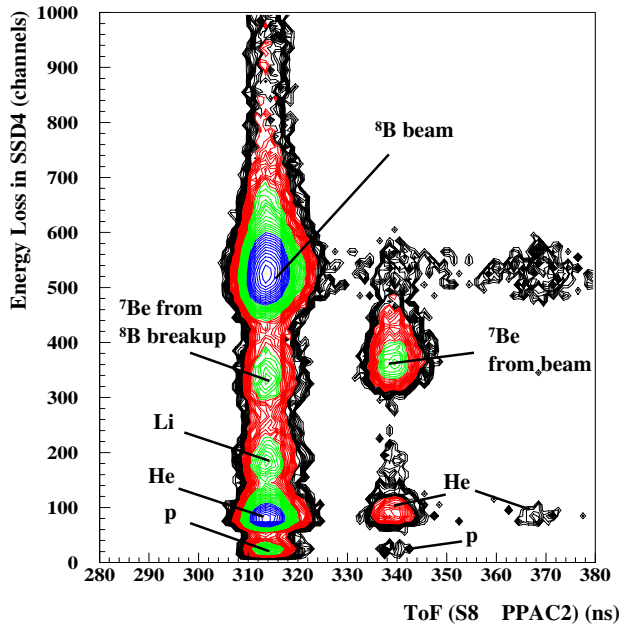} \vspace*{-1.0cm}
\caption{(Color online)
Identification of primary and secondary fragments by
energy loss and time of flight. } \label{de_tof}
\end{figure}

Typical $^8$B beam intensities in front of KaoS were $5 \times
10^4$ per 4 sec spill; the only contaminant consisted of about
20\% $^7$Be ions which could be identified event by event with the
help of a time-of-flight measurement. For this purpose a 3 mm
thick plastic scintillator detector was installed in the transfer
line between FRS and KaoS, about 85 m upstream from the breakup
target, to serve as a time-of-flight (ToF) start detector.
Positions and angles of the secondary beam incident on the Pb
breakup target were measured with the help of two parallel-plate
avalanche counters (PPAC) located at 308.5 cm and 71 cm upstream
from the target, respectively. The detectors, which were designed
and built at RIKEN \cite{Kum01}, had areas of $10 \times
10\,\text{cm}^2$ and allowed to track the incident $^8$B beam with
about 99\% efficiency and with position and angular resolutions of
1.3 mm and 1 mrad, respectively. In addition, they provided a ToF
stop signal with a resolution of 1.2 ns (FWHM). Fig.\ref{de_tof}
shows a two-dimensional plot of the ToF between the scintillator
detector and the second PPAC detector in front of the target.
One can see that a ToF measurement alone is sufficient to separate
the $^8$B beam from contaminants on an event-by-event basis.

\subsection{Detection of break-up fragments}

\begin{figure*}[bt]
\includegraphics[width=17.0cm]{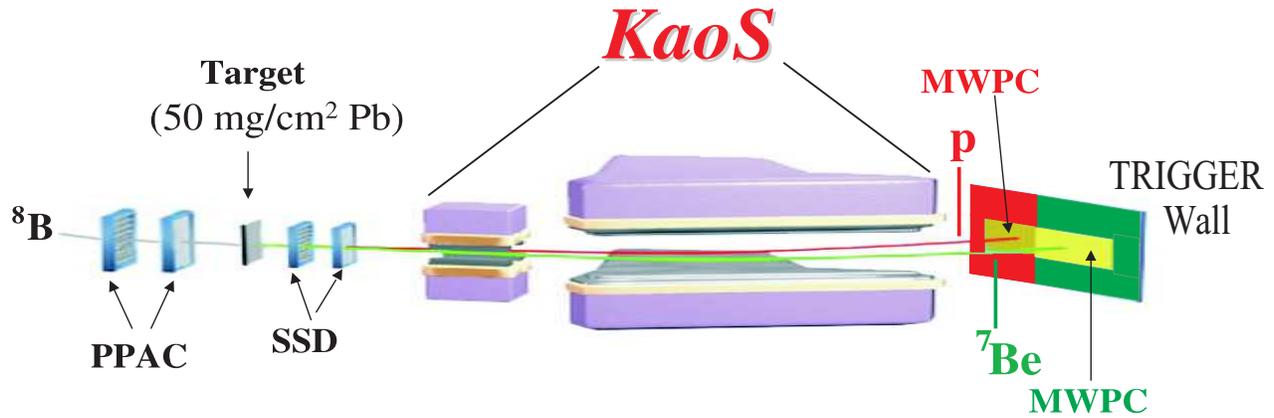}
\caption{(Color online)
Artist's view of the experimental setup. Shown
schematically are the beam-tracking detectors (PPAC) in front of
and the fragment-tracking Si strip detectors (SSD) behind the
Coulomb-breakup target. Proton and $^7$Be positions in the focal
plane of the KaoS magnetic spectrometer are determined by
large-area multi-wire chambers (MWPC) followed by a
scintillator-paddle wall for trigger purposes.} \label{kaos_setup}
\end{figure*}

A schematic view of the experimental setup to detect the breakup
of $^8$B in semi-complete kinematics (i.e. without detecting
coincident $\gamma$-rays) at the KaoS spectrometer at GSI is shown
in Fig.\ref{kaos_setup}. Apart from the PPAC tracking detectors
mentioned above, it consisted of (i) the $^{208}$Pb breakup
target; (ii) two pairs of Si strip detectors; (iii) the magnets of
the KaoS spectrometer; (iv) two large-area multi-wire proportional
chambers (MWPC); (v) a ToF wall serving as a trigger detector. The
individual components will be discussed in detail below.

\subsubsection{Fragment tracking: Si strip detectors}

\begin{figure}[bt]
\vspace{0.8cm}
\includegraphics[width=7.5cm]{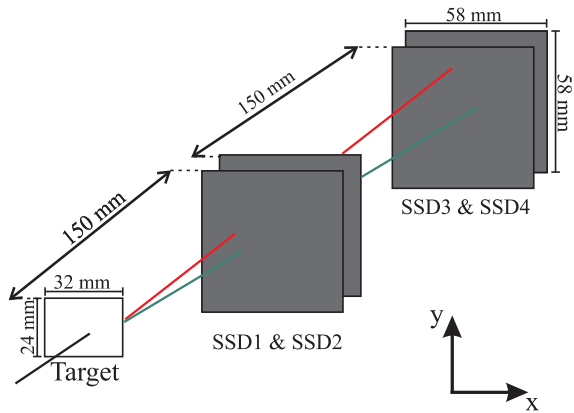}
\caption{(Color online)
Schematic view of the geometrical arrangment of the four
layers of single-sided Si strip detectors yielding the breakup
particles' trajectories directly after the target. }
\label{ssd_layout}
\end{figure}

Downstream from the Pb target (which consisted of 52 mg/cm$^2$
$^{208}$Pb enriched to 99.0$\pm$0.1\% and had an area of 24 mm in
height times 36 mm in width), the angles and positions as well as
the energy losses of the outgoing particles were measured with two
pairs of single-sided Si strip detectors (SSD). These detectors
(300 $\mu$m thick, 100 $\mu$m pitch) were located at distances of
about 15 cm and 30 cm downstream from the target.
Fig.\ref{ssd_layout} shows schematically the layout of the SSD
detector array.
The vacuum of the beamline housing the PPAC's, the target and the
SSD detectors was separated downstream from ambient air by a
stainless-steel window of 50 $\mu$m thickness.

\subsubsection{The KaoS magnetic spectrometer}

The KaoS magnetic spectrometer \cite{Sen93} consisted of a
large-aperture quadrupole and a horizontally-focussing dipole
magnet. The ratio between the smallest and largest momentum
accepted by KaoS amounted to about two, making KaoS an ideal
instrument to detect breakup of neutron-deficient nuclei into a
proton with $A/Z$=1 and an ion with $A/Z \approx 2$. Prior to our
measurement, the magnetic field of the KaoS dipole had been mapped
in three dimensions to obtain an empirical field map; this map was
then used to simulate the passage of charged particles through the
magnet using the code GEANT-3 \cite{geant}. To avoid multiple
scattering of the fragments in air, the chamber inside the
quadrupole and dipole magnets was filled with He gas at 1 bar
pressure, separated from the ambient air by thin He-tight foils.

\subsubsection{Fragment tracking: Multi-wire chambers and trigger detectors}

Behind the magnets, two large-area MWPC were installed as close to
the focal plane as possible. One chamber, with horizontal and
vertical dimensions of 60 cm and 40 cm, respectively, detected the
positions of protons behind KaoS. The other one, 120 cm wide and
60 cm high, was set to detect the $^8$B non-interacting beam and
the $^7$Be fragments. The separation of the position measurements
of protons and the heavy ions allowed to optimize each detector
voltage for optimum detection efficiency.

Behind the focal plane and parallel to it, a plastic-scintillator
wall with 30 elements (each 7 cm wide and 2 cm thick) was
installed and used for trigger purposes. The wall was subdivided
into two sections covering the respective MWPC in front of them.
Coincident signals in the left hand (proton) part and in the right
hand (ion) part of the wall indicated a breakup event (``breakup''
trigger). Singles hits in the right hand section were interpreted
as ``beam'' triggers and recorded with a down-scale factor of
1000.

\subsection{Monte-Carlo simulations}
Monte-Carlo simulations of the Coulomb breakup of $^8$B and the
detection of the breakup products are an indispensable part of the
present experiment. They were essential in planning the
experiment, helped to estimate the energy resolution and detection
efficiencies, and were instrumental in determining the proton and
$^7$Be momenta from the measured positions. As a tool for these
Monte-Carlo simulations the program package GEANT-3~\cite{geant}
was used.

The Monte-Carlo simulations started with an event generator that
simulated CD of $^8$B on $^{208}$Pb in first-order perturbation
theory or via a fully dynamical calculation by the theoretical
approaches mentioned above (subsect.II.B). Technically, the event
generator produced statistically-distributed ensembles of 500,000
CD-``events'' each that were used as input to a GEANT simulation
of the passage of each breakup particle through the Pb target, the
SSD detectors, the beamline exit window, the He-filled interior of
the magnets and the air behind KaoS before hitting the MWPC
volumes. At the target, the emittance of the $^8$B as measured
with the PPAC's was imposed, the momentum spread was assumed to be
the nominal FRS momentum acceptance, $\Delta p/p = \pm 1 \%$.

\begin{figure}[bt]
\includegraphics[width=7.5cm]{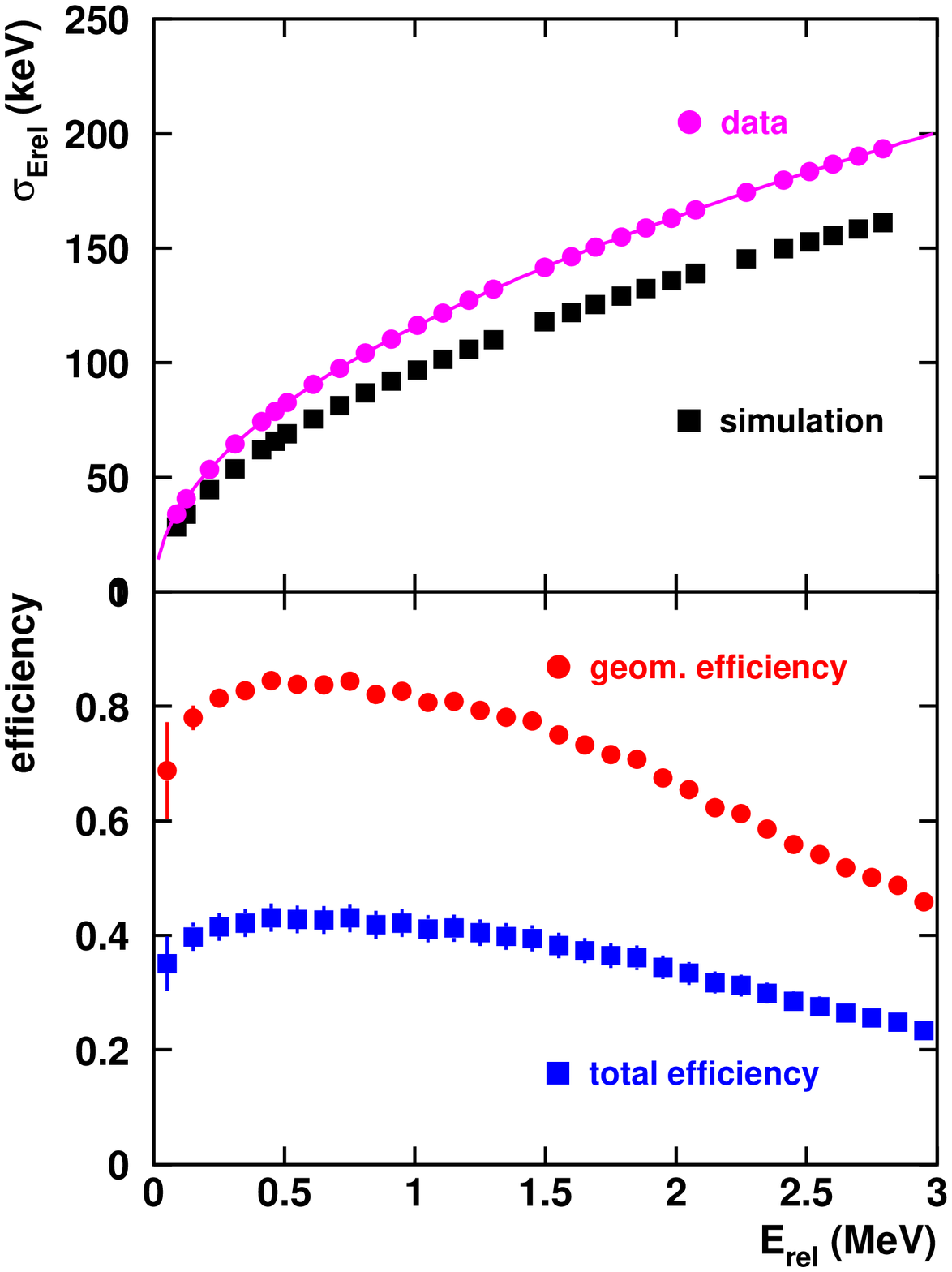}
\vspace*{-1.0cm}
\caption{(Color online)
Calculated properties of the experimental setup from
GEANT simulations. Top: $E_{rel}$ resolution (squares). The upper
set of points (circles) has been determined from the experimental
data. \\
Bottom: Geometrical efficiency (upper set of data points).
In addition, we show in the lower set of data points also the
total efficiency including all analysis conditions. }
\label{minv_effic}
\end{figure}

Momenta of each particle type (p,$^7$Be,$^8$B) were obtained from
two position measurements in the SSD and one position measurement
in the respective MWPC. To calculate each particle type's
momentum, a 36-term polynomial expression was derived; its
parameters were obtained in a GEANT simulation by sending
particles with known momenta (covering evenly the range of
relevant momenta) through the setup and fitting the momenta as a
function of the positions by varying the 36 polynomial parameters.
In a similar way, the invariant-mass resolution of the experiment
could be obtained by simulating breakup events of known invariant
mass and reconstructing this quantity from the simulated
positions. The top panel in Fig.\ref{minv_effic} shows the
$E_{rel}$ resolution ($1 \sigma$ width) as a function of the
p-$^7$Be relative energy, $E_{rel}$, as determined from the
simulation.

The efficiency of our setup at high $E_{rel}$ is mainly given by
the finite sizes of the SSD and MWPC detectors. Below the maximum
around 0.5 to 1 MeV, the efficiency drops due to overlap of the
proton- and $^7$Be hit patterns in the SSD leading to apparent
multiplicity 1 instead of 2. Numerical values of the efficiency
could be obtained by simulating the full set of 500,000 CD events
with and without the above conditions and plotting the ratios of
these numbers for different, evenly spaced $E_{rel}$ bins. This
distribution is shown in the lower panel of Fig.\ref{minv_effic}.
The upper set of data points (circles) is obtained by requiring
two separated p-Be hits inside all detector volumes. The lower set
of data points (squares) is obtained by taking into account the
intrinsic detector and trigger efficiencies and applying all
analysis conditions, see subsect.~IV.B below. It can be seen that
the major part of the $E_{rel}$ distribution is covered with high
total efficiency (about 30-40\%). It should be noted that this
curve is insufficient to correct measured data for efficiency: the
total efficiency is a multi-dimensional function of both the
original and the smeared-out (by the experimental resolution)
angles and momenta of both particles. Therefore, we pass the
theoretical ``events'' through the experimental filter and compare
the results to the same quantities derived from the data.

\section{Data reduction and results}

The experiment described in the present paper recorded events
from three different sources:\\
(i) breakup events originating in the Pb target;\\
(ii) down-scaled beam particles;\\
(iii) background from a variety of sources (e.g. cosmic rays).\\
Though event classes (i) and (ii) are mainly correlated with a
corresponding trigger type (``breakup'' trigger for class (i),
``beam''  trigger for class (ii)) we have checked if by chance the
trigger types and event classes were mixed in rare cases, and have
corrected for that.

In the following, we first show how the total number of incident
$^8$B projectiles is obtained from the ``beam'' trigger events. We
then explain how the breakup events originating in the Pb target
were identified.

\subsection{Total number of $^8$B projectiles}
The absolute number of $^8$B ions impinging on the $^{208}$Pb
breakup target needs to be known to determine absolute cross
sections. To this end, ``beam'' trigger events were analyzed to
select those that correspond to $^8$B in the $\Delta E$-ToF plot,
Fig.~\ref{de_tof}. A 3$\sigma$ window around the $^8$B energy-loss
peak in each SSD was chosen. To convert the integrated number in
this spectrum to the total number of incident $^8$B ions, the
down-scale factor of the ``beam'' trigger (10$^3$) and the
efficiencies for detecting $^8$B ions in the ToF detectors (3 mm
scintillator, PPAC detectors) as well as in the $\Delta E$ (SSD)
detectors have to be taken into account. These numbers were
derived from sets of linear equations containing the coincidence
count rates and the respective efficiencies. A small number
(0.48\%) of $^8$B ions was found for the ``breakup'' trigger
condition due to random-noise coincidences with the left
(``proton'') part of the plastic-scintillator wall. Together with
the total from the ``beam'' trigger condition we obtain a total of
$(4.15 \pm 0.03) 10^9$ $^8$B ions impinging on the breakup target.
\begin{figure}[bt]
\vspace{-0.5cm}
\includegraphics[width=9.0cm]{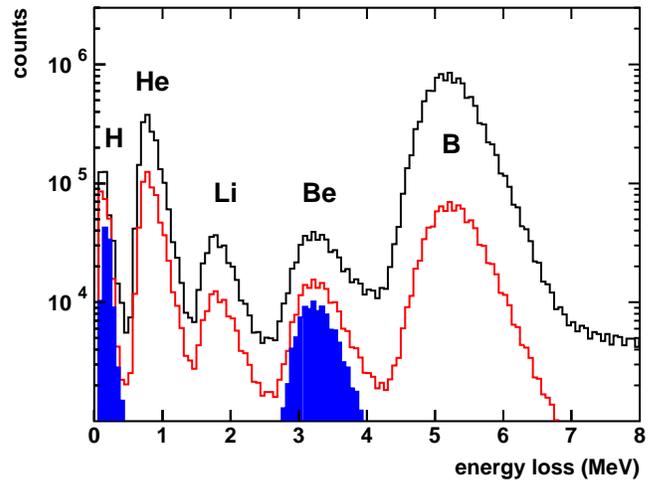}
\caption{(Color online)
Energy-deposition of the incident $^8$B ions and of the
breakup fragments in the third Si strip detector. The outermost
contour corresponds to all events. The intermediate thin contour
depicts events where an incident $^8$B particle is correlated with
multiplicity $m = 2$ in each SSD.  The innermost filled histograms
are obtained by requiring a $\Delta E$-cut on $^7$Be in each SSD
plus a p-$^7$Be vertex inside the target volume (see text). }
\label{de_ssd}
\end{figure}

\subsection{Identification and tracking of breakup products}

The coincident p and $^7$Be signals resulting from breakup in the
$^{208}$Pb target were identified among the class (i) events
(``breakup'' trigger) in several successive steps:
\begin{enumerate}
\item The $\Delta E$-ToF condition was applied to select only
incident $^8$B ions (see above). \item A multiplicity of $m \geq
2$ in each SSD was required. That meant that at least one empty
strip was found between two adjacent hit clusters. \item A
3$\sigma$-window around the $\Delta E$ peak corresponding to the
energy loss of $^7$Be in each SSD selected $^7$Be as one of the
reaction products. \item The coincident protons were found among
all events with $\approx 50$ keV $ < \Delta E < 500$ keV in each
SSD where the trajectories had a closest distance to the
coincident $^7$Be trajectory inside a volume given in x and y by
the size of the target ($\pm$18 mm in x- and $\pm$12 mm in
y-direction) and in z-direction (along the beam axis) of $\pm$25
mm around the target (located at $z=0$). The low-energy cutoff was
chosen individually for each detector; the number of protons below
this cutoff was estimated by fitting a Gaussian to the low-energy
tail of the Landau distribution.
\end{enumerate}
The inclusive $\Delta E$ spectra resulting from conditions 1 and 2
above are shown by the intermediate thin histogram in
Fig.\ref{de_ssd}, whereas conditions 3 and 4 lead to the innermost
(filled) histograms in Fig.\ref{de_ssd}. This procedure removed
all breakup events in layers of matter other than the target and
led to a practically background-free measurement.

The breakup protons loose only about 200 keV in the 300 $\mu$m
thick SSD. Nevertheless, after imposing the vertex condition, the
energy-deposition signals of protons in the SSD are clearly
resolved from noise.

\subsection{Invariant-mass reconstruction}

\subsubsection{Proton-ion opening angles}

The p-$^7$Be relative energy, $E_{rel}$, is derived from the total
energies, $E$, of the particles, their 3-momenta, $p$, and the
p-Be opening angle, $\theta_{17}$ according to
\begin{equation}
E_{rel} = \sqrt{(E_{Be}+E_p)^2 -p_{Be}^2 -p_p^2 -2 p_{Be} p_p
\cos(\theta_{17})}. \label{eq1}
\end{equation}
Whereas the proton and $^7$Be momenta can be obtained only from
the rather complicated momentum reconstruction procedure described
in the next subsection, the p-$^7$Be opening angle, $\theta_{17}$,
can be determined directly from constructing the vectors
connecting the breakup vertex in the target with the corresponding
hit positions in the SSD. Since protons fire only a single strip
in each SSD, their positions are given by the strip centroid, and
the variances of these positions - assuming that the hits are
evenly distributed over the strip width - by the the strip pitch,
100 $\mu$m, divided by $\sqrt{12}$. In contrast, the larger energy
deposits by the $^7$Be ions produce broader hit patterns in our
setup, with rather large fluctuations of the widths.

\begin{figure}[bt]
\vspace{-0.5cm}
\includegraphics[width=9cm]{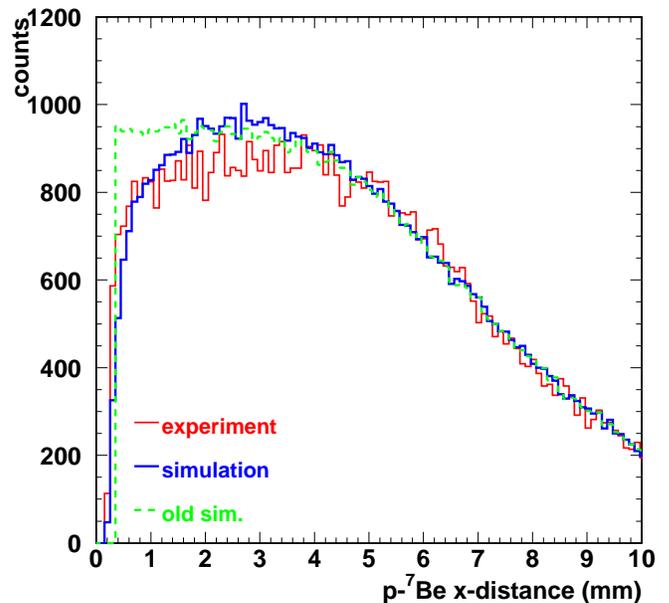}
\vspace*{-1.2cm}
\caption{(Color online)
Horizontal (x-) distances between proton and $^7$Be hits
in the first Si strip detector. The thin (red) histogram shows the
distribution of experimental distances, the thick (blue) one those
from the present GEANT simulation. The dashed (green) histogram
shows the GEANT simulation that was used to evaluate our previous
results \cite{Sch03}. } \label{p_Be_dist}
\end{figure}

To reconstruct a breakup event, the p and $^7$Be hits in each SSD
have to be separated by at least one empty strip. Since this
affects the efficiency for identifying a breakup event for low
$E_{rel}$, we have to make sure that the GEANT simulation
accurately reproduces this efficiency. This has been achieved by
introducing a weighting function in GEANT that gradually increases
the efficiency for detecting two separated hits from zero to one
over the appropriate distance for each detector so that
experimental and simulated distance distributions look alike. In
Fig.\ref{p_Be_dist} we plot the horizontal-distance distribution
between
proton and $^7$Be hits in the first SSD. One can observe that
experiment and simulation yield very similar distributions. It
should be emphasized that in our earlier data analysis a step
function was assumed for this efficiency that jumped from zero to
full efficiency at a fixed distance of 0.4 mm in each SSD. This is
visualized by the dashed histogram in Fig.\ref{p_Be_dist}; it
clearly shows that we overestimated the GEANT detection efficiency
for small $E_{rel}$ in our previous paper~\cite{Sch03}. As we will
show below, this leads to slightly larger cross sections at low
$E_{rel}$ compared to those of Ref.~\cite{Sch03}.

\begin{figure}[hb]
\vspace{-1.0cm}
\includegraphics[width=6cm]{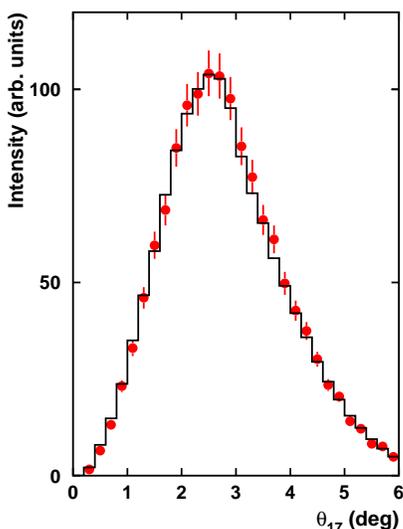}
\vspace*{-0.5cm}
\caption{(Color online)
Distribution of opening angles between proton and
$^7$Be, $\theta_{17}$. Data points are shown by the symbols, the
histogram indicates the GEANT simulation. } \label{theta17}
\end{figure}

The validity of this procedure can be checked immediately
by inspecting the
integral distribution of p-$^7$Be opening angles, $\theta_{17}$,
both in experiment and
in simulation. These distributions are shown in Fig.\ref{theta17}.
The agreement is excellent.

\subsubsection{Momentum reconstruction of the fragments}
As mentioned already above, momenta for each particle
(p,$^7$Be,$^8$B) where calculated from two position determinations
in front of KaoS (in the SSD) and from another position
determination behind KaoS (in the MWPC). These 6 coordinates were
converted to momenta using three sets of 36-term polynomial
expressions, one for each ion. By combining event by event the
longitudinal momenta of p and $^7$Be, one can check the accuracy
of the momentum reconstruction; its width is a measure how well
angular straggling effects are treated in the GEANT simulation.
The comparison shows that the simulated momentum widths are more
narrow by 20\% compared to experiment. Therefore, the simulated
$\sigma_{Erel}$ values shown in Fig.\ref{minv_effic} (squares)
where uniformly increased by 20\% (circles).

\subsection{Angular distributions}
\begin{figure}[bt]
\vspace{0.8cm}
\includegraphics[height=4cm]{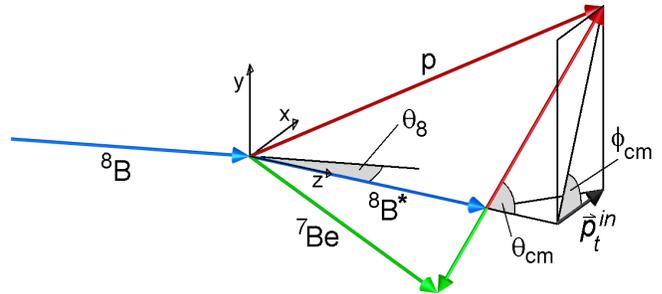}
\caption{(Color online)
Vector diagram showing the definitions of the angles
$\theta_{cm}$ and $\phi_{cm}$ as well as the proton in-plane
transverse momentum, $p_t^{in}$, in the frame of the $^{8}$B$^*$
system.} \label{vector_diagram}
\end{figure}

In the following we will present some angular distributions that
can be shown to be sensitive to an E2 amplitude in CD.
Fig.~\ref{vector_diagram} shows the coordinate systems used. With
$^8$B$^*$ we denote the momentum vector of the (excited) $^8$B
prior to breakup, as it is reconstructed from the measured proton
and $^7$Be momentum vectors. The angle $\theta_8$ is the
laboratory scattering angle of $^8$B$^*$ relative to the incoming
$^8$B beam. The polar angles, $\theta_{cm}$, and the azimuthal
angles, $\phi_{cm}$, of the breakup protons are measured in the
rest frame of the $^8$B* system. In the same way, one can
calculate the transverse proton momentum vector in the reaction
plane ($p_t^{in}$).

\subsubsection{Comparison to perturbation-theory calculations}

\begin{figure}[bt]
\includegraphics[width=8.0cm]{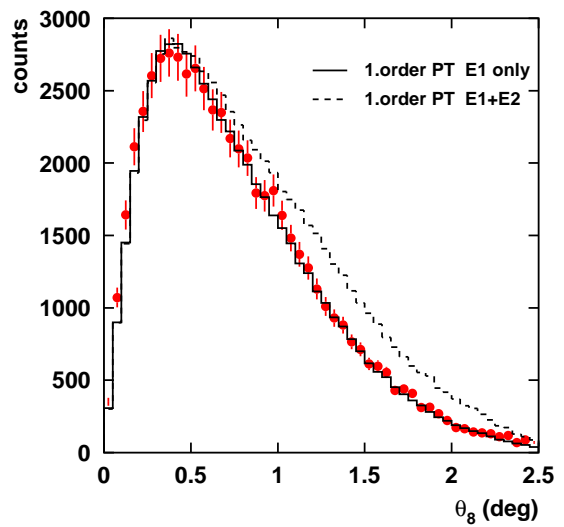} \vspace{-1.0cm}
\caption{(Color online)
Scattering angle $\theta_8$ of the excited $^8$B prior to
breakup, as reconstructed from the proton and $^7$Be vectors. The
full histogram has been calculated in first-order perturbation
theory assuming pure E1 multipolarity, the dashed one assuming
E1+E2 multipolarity.} \label{theta8}
\end{figure}

Fig.\ref{theta8} shows the $\theta_8$ distribution, in comparison
to two model calculations using first-order PT as discussed in
subsect.~II.B. The full histogram denoting pure E1 multipolarity
follows the data points very well, even to very large angles. The
dashed histogram, where both E1 and E2 with their full theoretical
strengths were assumed, deviates from the data points markedly for
$\theta_8$ values above about 0.7 degrees. Note that the
theoretical histograms were folded with the experimental response.
We conclude that even the $\theta_8$ distribution already
indicates E1 dominance, in a similar way as demonstrated in the
paper by Kikuchi \textit{et al.} \cite{Kik98}.

\begin{figure}[t]
\vspace*{-8mm}
\includegraphics[width=9.0cm]{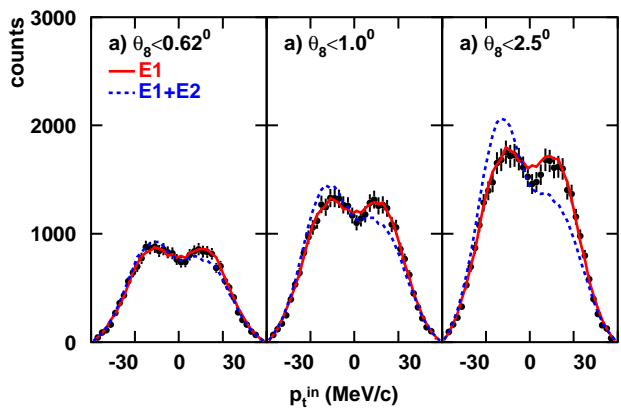}
\vspace*{-0.8cm}
\caption{(Color online) In-plane transverse momenta, $p_t^{in}$,
of the breakup protons for three different cuts in $\theta_8$. The
theoretical curves (full red lines: E1 multipolarity, dashed blue
lines: E1+E2 multipolarity) have been calculated in first-order
perturbation theory. They were normalized individually to the data
sets in each frame. } \label{ptin}
\end{figure}

We present in Fig.~\ref{ptin} the distribution
of $p_t^{in}$ for three different upper limits in $\theta_8$,
0.62$^{\circ}$, 1.0$^{\circ}$, and 2.5$^{\circ}$.
In classical Rutherford scattering, this corresponds to impact parameters
of 30 fm, 18.5 fm, and 7 fm, respectively.
Relative energies between p and $^7$Be
up to 1.5 MeV were selected.
The experimental data for all three $\theta_8$-cuts can be
reproduced well by a PT calculation that includes only E1
multipolarity (full histograms in Fig.~\ref{ptin}, the theoretical
curves were normalized individually to the data sets). If
E1-plus-E2 multipolarity is used in the PT calculation, the
different impact-parameter dependences of E1 and E2 multipolarity
lead to markedly different shapes for the different
$\theta_8$-cuts (dashed histograms in Fig.\ref{ptin}). In
particular for large values of $\theta_8$, the latter
distributions show a large asymmetry with respect to $p_{t}^{in} =
0$ that is in clear disagreement with our data points.

By comparing Fig.\ref{ptin} with similar plots in our earlier
Letter (Fig.2 of Ref.\cite{Sch03}) one can see the improvement in
the GEANT simulation which was achieved by the modified
prescription for the p-Be hit resolution (see subsect.~IV.3.1).
The dips near $p_t^{in} \approx 0$ in the theoretical
distributions are now much closer to the experimental ones (though
small residual discrepancies are still visible in the rightmost
panel).

\begin{figure}[bt]
\vspace{-0.8cm}
\includegraphics[width=9.0cm]{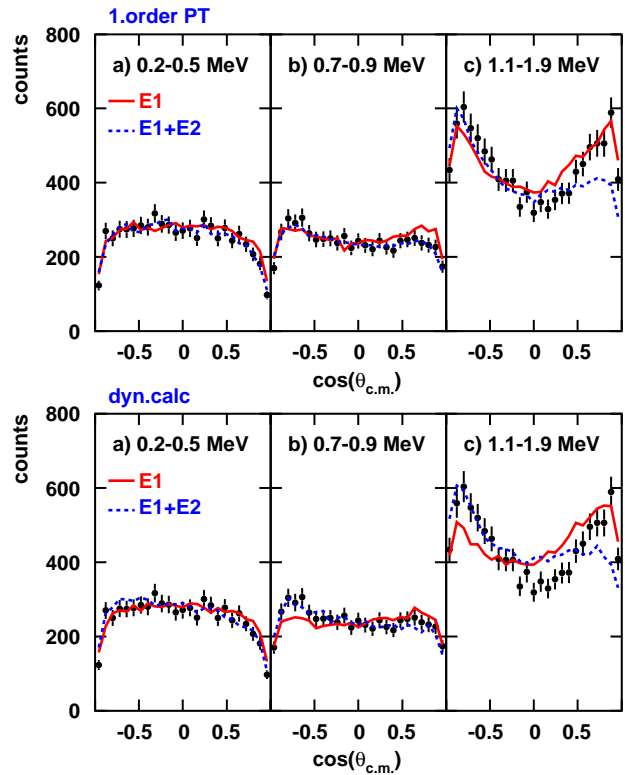}
\vspace*{-1.0cm}
\caption{(Color online)
Top: Experimental proton polar angle ($\theta_{c.m.}$)
distributions for three different bins of the p-$^7$Be relative
energy, $E_{rel}$. The full red curves denote a first-order
perturbation-theory calculation for E1 multipolarity, the dashed
blue ones for E1+E2. All theoretical curves were individually
normalized to the data points in each frame. Bottom: the same data
compared to dynamical calculations, again for E1 (full curves) and
for E1+E2 (dashed curves) multipolarities (see text for details).
} \label{theta}
\end{figure}

Fig.\ref{theta} depicts the experimental
$\theta_{cm}$ distributions for three different $E_{rel}$ bins,
as indicated in the figure. A ``safe'' $\theta_8$ limit of 1$^{\circ}$
was chosen.
As expected, these distributions are mostly isotropic at low
$E_{rel}$ (indicative of $s$-waves) and become increasingly
anisotropic for larger values (contributions from higher orbital
angular momenta). As in Fig.\ref{ptin}, also for the $\theta_{cm}$
distributions the calculations for pure E1 multipolarity fit all
spectra well; inclusion of an E2 component may lead to a slightly
better fit at low $E_{rel}$, but diverges clearly for the
large-$E_{rel}$ bin where E2 should play a major role. The
calculations with a dynamical model will be discussed below.

\subsubsection{Comparison to dynamical calculations}

As mentioned above, Esbensen \textit{et al.}~\cite{Esb04,Esb05}
suggested that dynamical calculations are required to properly
describe CD and to evaluate $S_{17}$ from the measured CD cross
sections. A sensitive test if such a theory describes the
experimental data better than first-order PT calculations is given
by comparing the dynamical predictions (using the model described
in subsect.~II.B.2) to the same angular distributions (bottom part
in Fig.\ref{theta}). In all three frames shown, our E1-only
dynamical calculations do not agree well with the data points.
Dynamical calculations with E1+E2 seem to introduce a slight
improvement as long as the effect of E2 multipolarity is small,
but a major discrepancy shows up when E2 should have a stronger
influence (rightmost lower panel in Fig.\ref{theta}).

In general, one would expect that the more complete description of
the Coulomb breakup within the semiclassical dynamical approach
leads to a better agreement with the experimental data than the
simpler perturbative treatment. However, in the dynamical
calculation more model parameters that are not really constrained
have to be specified than in the first-order approach. E.g., the
results for the angular distribution in the dynamical calculation
depend crucially on the assumed E2 strength, i.e.\ there is a
considerable model dependence. With sufficicently precise
experimental data it would be possible to determine this strength
in a fitting procedure but this requires extensive calculations.
Additionally, one has to keep in mind that a full quantal
treatment of the breakup process could lead to different results.

We conclude that within the limits of our experimental conditions
the simplest model (first-order PT with E1 multipolarity only)
still gives the best agreement with the measured center-of-mass
proton angular distributions. This is in line with conclusions
drawn by Kikuchi \textit{et al.}~\cite{Kik98} and by Iwasa
\textit{et al.}~\cite{Iwa99} from their respective $\theta_8$
distributions (which are, however, less sensitive to a small E2
component than the present angular correlations). Our findings
contradict the conclusions of Davids \textit{et al.}~\cite{Dav01}
that a substantial E2 cross section has to be subtracted from the
total measured CD cross section.

What remains to be done is to find a physical explanation for the
small E2 strength compared to the model calculations (both, the
potential model and the cluster model, predict almost equal
$S_{17}^{E2}$ values). At the same time, one has to find a
different way to explain the asymmetries found in inclusive
longitudinal-momentum distributions~\cite{Dav01} and attributed
either to a quenched~\cite{Dav01} or
enhanced~\cite{Dav03,Mor02,Sum05} E2 strength relative to the
respective model calculations.

\subsection{Energy-differential dissociation yields}

\begin{figure}[bt]
\includegraphics[width=9.0cm]{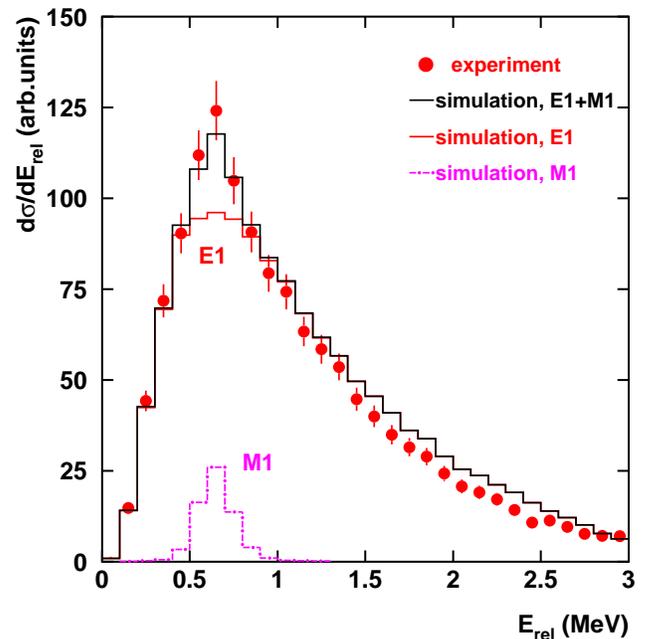}
\vspace*{-1.0cm}
\caption{(Color online)
Energy-differential Coulomb-dissociation yields for
equal-sized $E_{rel}$ bins of 100 keV each. The thick outermost
histogram results from our GEANT simulation including E1 and M1
multipolarity, scaled by a factor of 0.82. The thin(dash-dotted)
histograms show the separate contributions from E1(M1)
multipolarity. } \label{dsde}
\end{figure}

The measured momentum vectors of the outgoing p and $^7$Be
particles allowed to calculate $E_{rel}$ according to Eq.(1), from
which we have constructed the energy-differential dissociation
yields of the excited $^8$B$^*$ system prior to breakup
(Fig.\ref{dsde}).
In line with our findings of a negligible E2 contribution
discussed above, we compare this spectrum to a simulated one that
contains contributions from E1 and M1 multipolarities only. The
latter contribution was calculated using the M1 resonance
parameters as determined by Filippone \textit{et
al.}~\cite{Fil83}. As expected, M1 contributes only in a narrow
energy range around the peak of the spectrum. In plotting
Fig.\ref{dsde}, we have restricted the Rutherford scattering
angles $\theta_8$ to values below 1.0$^{\circ}$ to ensure both
dominance of CD and reduction of the effect of any possible E2
contribution.

It should be noted that in CD, starting from the $^8$B ground
state, both the ground state and the first excited state at 429
keV in $^7$Be can be observed as a result from first-order E1
excitation. The relative amount of these contributions to the CD
are determined by the relative spectroscopic factors of the two
${}^{7}$Be states  in the ${}^{8}$B ground state and the different
photon spectra due to the different excitation energies. This
component, which can be traced experimentally by observing the
coincident 429 keV $\gamma$-rays, needs to be subtracted before
calculating $S_{17}$ from differential CD cross sections.
Numerical values for this branching have been kindly provided by
Kikuchi \textit{et al.}~\cite{Kik98} and were scaled to the
present bombarding energy using Weizs\"acker-Williams theory.

Since the shape of the theoretical $d\sigma/dE$ distribution is
better defined than its absolute magnitude, we have normalized
both distributions to each other; the resulting scaling factor is
$f=0.82$. With this renormalization, the experimental and
simulated distributions agree rather well (Fig.~\ref{dsde}). Small
deviations between the data points and the histogram indicate
discrepancies between the assumed $S_{17}$ factor from our
potential model and the true one, as will be discussed in the next
section.

\section{The astrophysical $S_{17}$ factor}
\begin{figure}[bt]
\includegraphics[width=9.0cm]{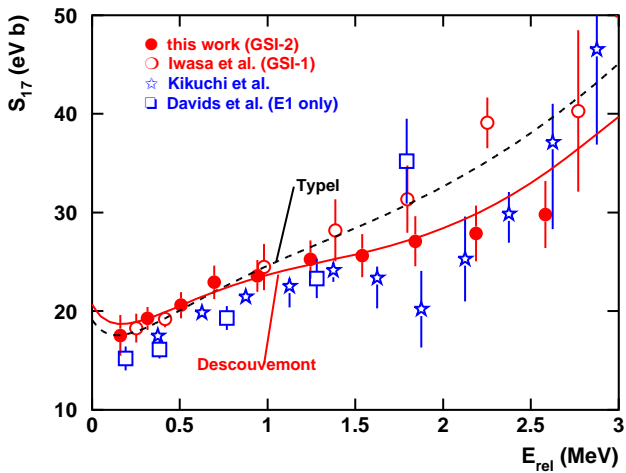}
\caption{(Color online)
Comparison between $S_{17}$ values from
Coulomb-dissociation experiments. The full (open) circles indicate
the present (previous) GSI CD experiment labelled GSI-1 (GSI-2).
Open stars depict Ref.\protect\cite{Kik98}, open squares
Ref.\protect\cite{Dav01} (E2 contribution subtracted). The
theoretical curves are described in the text. } \label{s17_cd}
\end{figure}

The measured quantity in CD of $^8$B is the distribution of
energy-differential cross sections, Fig.\ref{dsde}. This
distribution is related to $S_{17}$ via a theoretical model. We
assume that at the high incident energy used in our experiment and
for the low $Q$-value of the reaction, first-order perturbation
theory is adequate to describe Coulomb dissociation. This has been
investigated in detail in Ref.~\cite{Typ02}. In analyzing our
results, we also assume that the GEANT simulations describe all
experimental effects quantitatively, in particular the feeding of
neighboring bins due to the relatively bad $E_{rel}$ resolution.
We have verified this assumption by comparing data and simulations
for several raw observables, e.g. the $\theta_{17}$ distribution
of Fig.\ref{theta17} or the $\theta_8$ distribution of
Fig.\ref{theta8}. Based on the good agreement, we conclude that
any remaining discrepancies between the two histograms in
Fig.\ref{dsde} can be attributed to a deviation of the true E1
$S_{17}$ factor from the one used in our simulation. Thus, the
true $S_{17}$ factor for each bin was obtained by multiplying the
theoretical one (averaged over this bin width) by the ratio of
observed and simulated counts. The bins were chosen in accordance
with the $E_{rel}$ resolution (Fig.\ref{minv_effic}) to be roughly
one FWHM wide, i.e. between 0.2 and 0.3 MeV. The resulting
$S_{17}$ factors as a function of $E_{rel}$ are visualized in
Fig.~\ref{s17_cd} and listed in Table \ref{s17_table}.
\begin{table}
\caption{\label{s17_table}Numerical values of $S_{17}$ (in eV b)
as a function of $E_{rel}$ (in MeV). The 1-sigma errors include
all $E_{rel}$-dependent terms. A common systematic error of 5.6\%
has to be added in quadrature to each data point (see text).}
\begin{ruledtabular}
\begin{tabular}{ccc}
$E_{rel}$(\text{MeV})& $S_{17}$\text{(eV b)} & $\sigma_{S17}$ \text{(eV b)}\\
\hline
0.160 & 17.5 & 2.1\\
0.316 & 19.3 & 1.2\\
0.507 & 20.6 & 1.3\\
0.695 & 22.9 & 1.7\\
0.942 & 23.6 & 1.6\\
1.244 & 25.2 & 1.9\\
1.540 & 25.6 & 2.2\\
1.841 & 27.1 & 2.5\\
2.187 & 27.9 & 2.8\\
2.582 & 29.8 & 3.4\\
2.988 & 56.8 & 7.1\\
\end{tabular}
\end{ruledtabular}
\end{table}

The error bars shown in  Fig.~\ref{s17_cd} and listed in Table
\ref{s17_table} contain all $E_{rel}$-dependent terms, resulting
from counting statistics, from the error of the geometrical
efficiency as determined by the GEANT simulations, and from the
error of the feeding of the excited state in $^7$Be. In addition,
uncertainties in determining $E_{rel}$ and $\theta_8$ are
included. An $E_{rel}$-independent systematic error of 5.6\% has
to be added for all data points, reflecting an estimated error of
the dead-time correction (0.6\%), of the number of incident $^8$B
projectiles (1.4\%), and of the analysis (5.4\%).  The analysis
error consists of the combined errors related with choosing the
appropriate gates to identify a $^7$Be fragment in the $\Delta E -
ToF$ spectra (1.8\%) and to identify a proton via the vertex
reconstruction (5.1\%). The latter contribution reflects the
uncertainty in choosing the low-energy cutoff in the
proton-$\Delta E$ spectra to remove the noise, which at the same
time leads to the loss of some real proton events. More details
can be found in Ref.~\cite{Sch02}.

\subsection{Comparison with other CD experiments}
Fig.\ref{s17_cd} shows the astrophysical $S_{17}$ factors as
deduced from the three other CD experiments published so far
\cite{Kik98,Iwa99,Dav01} (the data of Ref.\cite{Dav01} represent
their E1-$S_{17}$ factors after subtraction of the
E2-contribution). The CD $S_{17}$ factors are in reasonable
agreement with each other, though both the Kikuchi \textit{et
al.}~\cite{Kik98} and the Davids \textit{et al.}~\cite{Dav01} data
are systematically lower. We note that also our earlier CD
experiment~\cite{Iwa99} and the present one are in good agreement
up to $E_{rel} \approx 1.5$ MeV, marked discrepancies occur only
at higher $E_{rel}$ values. Compared to our previous results given
in Ref.~\cite{Sch03}, the lowest three data points have been
increased by 6.7\%, 10\%, and 5.8\%, respectively. The remaining
data points remain largely unaffected. As a consequence, the slope
of our $S_{17}$ factors as a function of $E_{rel}$ becomes smaller
and fits much better than previously to the energy dependence of
Descouvemont's cluster model; we will discuss this aspect in more
detail below.

In the above comparison with other published CD results, we have
plotted $S_{17}$ values as deduced by the authors in their
analyses. Other evaluations of the same data sets may lead to
different results. As an example, we quote the recent re-analysis
of the energy- and angle-differential cross sections, $d\sigma/dE$
and $d\sigma/d\theta_8$, of the RIKEN-2 experiment~\cite{Kik98} by
Ogata \textit{et al.}~\cite{Oga05}. The former authors deduced a
zero-energy factor of $S_{17}(0)=18.9 \pm 1.8$ eV b based on
first-order perturbation theory. Ogata \textit{et al.} obtain
$S_{17}(0)=21.4^{+2.0}_{-1.9}$ eV b from the same experimental
data, by taking into account interference of nuclear, E1 and E2
contributions and higher-order processes.

\subsection{Comparison with direct-capture experiments}
\begin{figure}[bt]
\includegraphics[width=9.0cm]{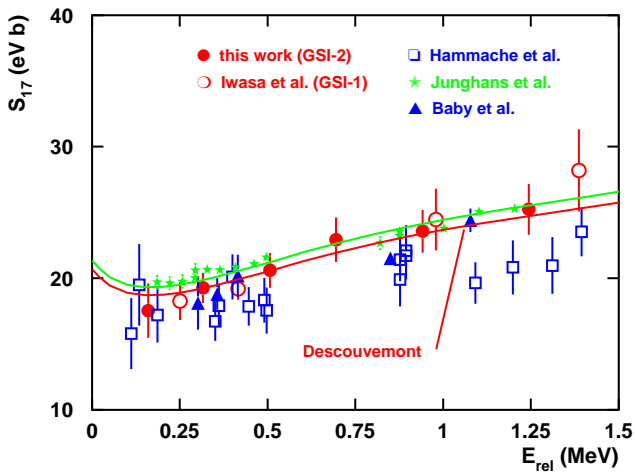}
\caption{(Color online)
$S_{17}$ from this work in comparison with the
(p,$\gamma$) experiments of Ref.~\protect\cite{Ham01} (squares),
Ref.~\protect\cite{Jun03} (stars), and Ref.~\protect\cite{Bab03}
(triangles). The latter data were corrected for the
contribution of the M1 resonance by the authors. The theoretical
curves are from Descouvemont~\protect\cite{Des04} and have been
fitted to the Seattle data (upper curve, Ref.~\cite{Jun03}) and
the present data (lower curve), respectively. See text for more
details.} \label{s17_cd_pgamma}
\end{figure}

 Fig.~\ref{s17_cd_pgamma} compares our data
to those of the recent $^7$Be(p,$\gamma$)$^8$B measurements where
the authors have subtracted the contribution from the M1 resonance
(Refs.~\cite{Ham01,Jun03,Bab03}). (Since we do not intend to dwell on
discrepancies among the results from direct-proton-capture
experiments, we restrict ourselves to the three data sets shown).
With the modifications of the lowest-$E_{rel}$ data points
discussed above, our dataset follows now closely the (p,$\gamma$)
data of Junghans \textit{et al.}~\cite{Jun03} over their {\em
entire} energy range. Previously~\cite{Sch03} we noted good
agreement with Refs.\cite{Jun03,Bab03} only for the data points
above the $M1$ resonance. This solves partly a puzzle that
Junghans \textit{et al.} claim to have observed, a systematic
discrepancy between the slope of the CD $S_{17}$ factors and those
from direct p-capture experiments. It also removes the
experimental basis for recommendations by Esbensen \textit{et
al.}~\cite{Esb05} to modify the deduced slope of $S_{17}(E_{rel})$
on the basis of a fully dynamical calculation.

\subsection{Extrapolation to zero relative energy}
To extrapolate to zero energy, all recent (p,$\gamma$) experiments
have chosen the cluster model of Descouvemont and
Baye~\cite{Des94}. As mentioned above, Descouvemont~\cite{Des04}
has recently refined the cluster-model description of $^8$B (we
refer to this model below as D04). In this refined approach, the
curve resulting from the Minnesota force (MN) is closer to the
experimental data and has been used in Fig.\ref{s17_cd_pgamma} to
fit both the Seattle data~\cite{Jun03} and our present results
over the energy range up to $E_{rel} = 1.5$ MeV. The fits yield
practically identical results within their respective errors. The
D04 normalization factor for our data set is $0.837 \pm 0.013$
with a reduced $\chi^2$ of 0.40. Note that Descouvemont has
investigated the error introduced by scaling the $S$ factor and
found it negligible~\cite{Des04}.

Our previous data set~\cite{Sch03} was found to be best compatible
with the potential-model calculation of Typel as discussed in
subsect.~II.B. of the present paper or in Davids and
Typel~\cite{Dav03} (referred to below as DT03). It is obvious that
with the modified low-energy data points of the present paper, the
agreement with this model is less satisfactory. The black dashed
curve in Fig.\ref{s17_cd} visualizes a fit of this theory to our
data. Though from a purely statistical point of view the fit with
the DT03 curve is acceptable, we prefer to describe our data with
the D04 theory for the following reasons:
\begin{enumerate}
\item The two-cluster structure of $^7$Be is related to its
intrinsic deformation, so that a model of $^8$B based on this
feature should be more realistic; \item the $S_{17}$ energy
dependence from all modern direct p-capture experiments can be
described consistently and fitted with high confidence with the
cluster model, thus corroborating the above conjecture; \item D04
allows to fit our lower 4 to 9 data points with practically equal
results; i.e. the scaling factor does not depend on the fit range.
In contrast, the DT03 scaling factor changes continuously with
increasing fit range, reflecting the different shape of the curve;
\item using D04, we find $\chi^2_{red} < 1$ for a fit range up to
2 MeV; the DT03 fit yields $\chi^2_{red}
> 1$ already if the fit range is extended above 1.3 MeV.
\end{enumerate}

When we fit our lowest 8 data points, up to $E_{rel}$ = 2 MeV, to
the D04 model, we obtain $S_{17}(0) = 20.6  \pm 0.8 $ eV b. The
same result within error bars is obtained if we use any number of
data points between four and eight. As mentioned above, a
systematic error of 5.6\% has to be added, yielding  $S_{17}(0) =
20.6 \pm 0.8 (stat) \pm 1.2 (syst)$ eV b. Not included in these
numbers is the theoretical uncertainty given by
Descouvemont~\cite{Des04} as about 5-10\%  depending on the
relative energy.

This result overlaps perfectly with a fit of D04 to the full data
set of Junghans \textit{et al.}~\cite{Jun03} which gives
$S_{17}(0) = 21.2  \pm 0.5 $ eV b. A fit of the Baby \textit{et
al.} (p,$\gamma$) data to the D04 model yields a very similar
result, $S_{17}(0)=19.8 \pm 1.0$ eV b. When fitting D04 to the
Hammache \textit{et al.}~\cite{Ham01} data set, we obtain a
smaller central value of $S_{17}(0)=18.4\pm 1.7$ eV b, but the
error bar still overlaps with ours.

\section{Conclusions}

We conclude that at sufficiently high incident energy, a
high-resolution exclusive Coulomb-dissociation experiment can
provide a rather precise value for the low-energy
$^7$Be(p,$\gamma$)$^8$B cross section. Among other conditions to
be fulfilled, the efficiency of the method as a function of
proton-$^7$Be relative energy has to be modelled precisely. By
setting tight constraints to the scattering angle $\theta_8$ and
analyzing proton-$^7$Be angular correlations, a significant
contribution from E2 multipolarity could be excluded. Compared to
our first study of $^8$B Coulomb dissociation~\cite{Iwa99}, we
could base this conclusion on carefully measured angular
distributions. In contrast to our earlier
publication~\cite{Sch03}, our re-analyzed results for the
astrophysical $S_{17}$ factor follow closely the energy dependence
as predicted by the refined cluster-model description of
Descouvemont~\cite{Des04}. This finding is in line with the most
recent measurements of the $^7$Be(p,$\gamma$)$^8$B reaction. The
combined statistical and systematic errors of our fit value for
$S_{17}(0)$ amounts to 6.6\%; a similar error contribution of
about 5\% comes from the model uncertainty~\cite{Des04}.

\acknowledgments
The authors wish to thank K.-H.~Behr, K.~Burkard, and
A.~Br{\"u}nle for technical assistance. Vivid discussions with
B.~Davids, P.~Descouvemont, M.~Hass, and A.~Junghans are
gratefully acknowledged.

\end{document}